\documentclass[%
reprint,
superscriptaddress,
floatfix,
amsmath,
amssymb,
aps]{revtex4-2}

\usepackage{graphicx}
\usepackage{rotating}
\usepackage{amsmath}
\usepackage{bbm}
\usepackage{subfigure}
\usepackage{color}
\usepackage{braket}
\usepackage{tikz}
\usepackage{hyperref}
\hypersetup{colorlinks, 
	linkcolor={blue!75!black!80!yellow},
	citecolor={blue!75!black!80!yellow}, 
	urlcolor={blue!75!black!80!yellow}
	}

\makeatletter \renewcommand\@make@capt@title[2]{%
\@ifx@empty\float@link{\@firstofone}{\expandafter\href\expandafter{\float@link}}%
\sffamily{\textbf{#1}}\@caption@fignum@sep#2 }

\begin{document}

\author{Fabijan Pavo\v{s}evi\'{c}$^{*,}$}
\affiliation{Department of Chemistry, Yale University, 225 Prospect Street, New Haven, Connecticut, 06520, USA}
\email{fpavosevic@gmail.com}
\author{Johannes Flick}
\affiliation{Center for Computational Quantum Physics, Flatiron Institute, 162 5th Ave., New York, 10010  NY,  USA}

\title[]
  {Polaritonic Unitary Coupled Cluster 
  for Quantum Computations}






\begin{abstract}

In the field of polaritonic chemistry, strong light-matter interactions are used to alter a chemical reaction inside an optical cavity. To explain and understand these processes, the development of reliable theoretical models is essential. While traditional methods have to balance accuracy and system size, new developments in quantum computing, in particular the Variational Quantum Eigensolver (VQE), offer a path for an accurate solution of the electronic Schrödinger equation with the promise of polynomial scaling and eventual quasi-exact solutions on currently available quantum devices. In this work, we combine these two fields. In particular, we introduce the quantum electrodynamics unitary coupled cluster (QED-UCC) method combined with the VQE algorithm, as well as the quantum electrodynamics equation-of-motion (QED-EOM) method formulated in the qubit basis that allows an accurate calculation of the ground-state and the excited-state properties of strongly coupled light-matter systems on a quantum computer. The accuracy and performance of the developed methods is tested for a H$_4$ molecule inside an optical cavity in a regime where strong electronic correlations become significant. For the first time, we explicitly include two photon effects from first principles. We show that the developed methods are in excellent agreement with the exact reference results and can outperform their traditional counterparts. The work presented here sets the stage for future developments of polaritonic quantum chemistry methods suitable for both classical and quantum computers.

\end{abstract}

\maketitle

\section{Introduction}
Recent experimental advances in the strong coupling regime have demonstrated the possibility to create light-matter quasiparticles in optical cavities, which can display severely different physical and chemical properties than their constituents~\cite{ebbesen2016hybrid,ruggenthaler2018quantum,FlickRiveraNarang2018}. These effects are utilized in the new field of polaritonic chemistry that studies the strong coupling between the light field and a molecular system and its ability to alter chemical reactions~\cite{lather2019cavity,campos2019resonant, thomas2016ground, thomas2019tilting,li2020resonance,Munkhbat2018} and other processes~\cite{xiang2020intermolecular,polak2020manipulating}. 
So far different theoretical methods have been developed to describe such systems, either using $\it ab$ $\it initio$ methods~\cite{tokatly2013time,ruggenthaler2014quantum} capable of describing the electronic structure accurately, but in practice limited to single and few molecule systems, or using reduced effective models~\cite{hage2017,feist2015} also valid in the collective strong coupling limit, but simplifying part of the electronic structure. Thus, the development of predictive $\it ab$ $\it initio$ theoretical models that are capable of describing the electronic structure accurately also in the collective strong coupling regime involving many molecules remains essential for the fundamental understanding and design of processes involved in polaritonic chemistry.

In quantum chemistry, among various approaches, the coupled cluster (CC) method offers a systematic way for obtaining a solution of the time-independent electronic Schrödinger equation~\cite{bartlett2007coupled,shavitt2009many}. In this method, the correlation effects between quantum particles are included via the exponentiated excitation cluster operator, and truncation of the cluster operator at a certain excitation rank establishes the hierarchy of the CC methods. Even at a low truncation rank, such as the coupled cluster method with single and double excitations and a perturbative treatment of triple excitations [CCSD(T)], the CC method provides a sufficient accuracy for many interesting chemical phenomena at a polynomial computational cost~\cite{raghavachari1989fifth}.

Along those lines, to study strongly light-matter coupled systems from first principles, recently the quantum electrodynamics coupled cluster (QED-CC) method~\cite{haugland2020coupled,mordovina2020polaritonic} has been introduced and is an extension of the CC method where now also photons in addition to electrons are treated quantum mechanically. We note that the QED-CC approach is naturally closely related to the ep-CC coupled cluster method, where electrons and harmonic phonons are treated quantum mechanically~\cite{white2020coupled}. For the QED-CC method, currently the only available implementation is the QED-CCSD-1 method that includes the coupled single and double electronic excitations and single photon excitations~\cite{haugland2020coupled}. It inherits all favorable properties of the CC method such as the high accuracy, as demonstrated in studies of noncovalent interactions~\cite{haugland2021intermolecular} and ionization energies~\cite{deprince2021cavity} in optical cavities, size-consistency, size-extensivity, and the polynomial computational cost~\cite{bartlett2007coupled,shavitt2009many,haugland2020coupled}. The last feature was gained at expense that this method, as well as its conventional electronic structure counterpart, is nonvariational, and the calculated ground state energy is not necessarily an upper bound of the exact ground state energy. This property can lead to the catastrophic behaviour in certain regimes where strong electronic correlation prevails~\cite{chan2004state}. To alleviate this issue, it is beneficial to design the variational variants of the CC method and its extensions where both electrons and photons are treated quantum mechanically. 
One such approach is the unitary coupled cluster (UCC) method~\cite{kutzelnigg1982quantum,bartlett1989alternative}. Although the UCC method is capable of mitigating this problem, it does so at a computational cost that scales exponentially with the system size even in a truncated form  on a classical computer; thus effectively prohibiting its application to larger system sizes.

An alternative route for solving the electronic Schrödinger equation is offered by recent developments in the field of quantum computing. Here, various approaches for achieving this goal have been proposed over the last two decades~\cite{cao2019quantum,bauer2020quantum,head2020quantum,bharti2021noisy}. Particularly successful and suitable for the currently available noisy intermediate-scale quantum (NISQ) devices~\cite{preskill2018quantum}, is the Variational Quantum Eigensolver (VQE) algorithm~\cite{peruzzo2014variational}. The VQE algorithm is a hybrid quantum-classical algorithm in which the preparation of the parametrized wave function ansatz and the measurement of the ground state energy is performed on a quantum device, whereas parameters that minimize the ground state energy are optimized using a classical algorithm on a classical computer. Because of its attractive features inherited from the CC methods as well as being unitary, the unitary coupled cluster with singles and doubles (UCCSD) method was used as the wave function ansatz in the original proposal of the VQE algorithm~\cite{peruzzo2014variational}. It has been repeatedly demonstrated that the VQE algorithm in tandem with the UCCSD ansatz can accurately predict ground-state~\cite{o2016scalable,romero2018strategies,lee2018generalized,grimsley2019adaptive,grimsley2019trotterized,kottmann2021feasible} and excited-state~\cite{lee2018generalized,mcclean2017hybrid,ollitrault2020quantum} properties of various molecular systems with polynomial costs on quantum computers with the potential to be applicable to large system sizes. 

\begin{figure*}[ht]
  \centering
  \includegraphics[width=6.5in]{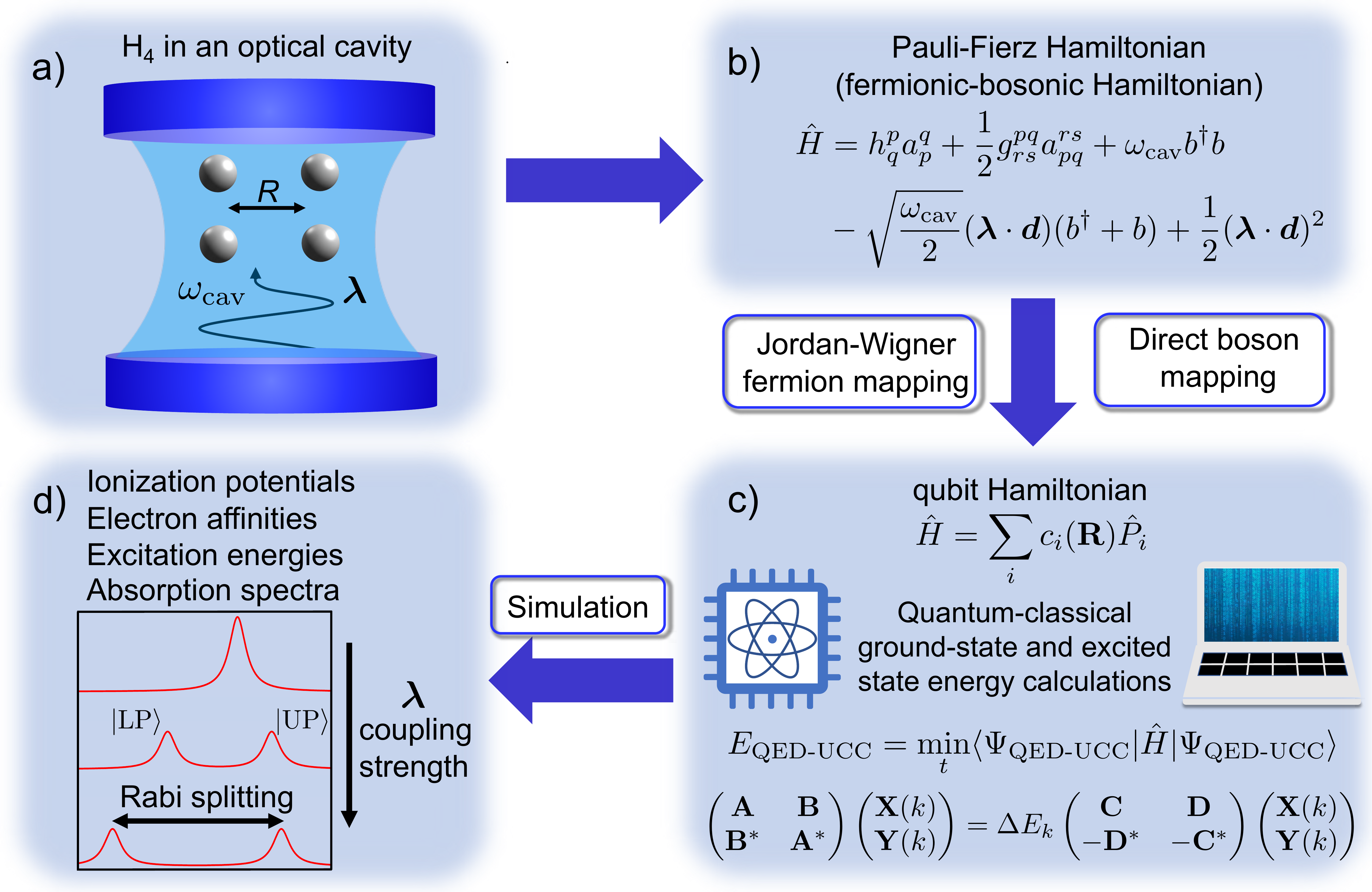}
  \caption{Schematic depiction of the steps performed in this work. (a) Illustration of the H$_4$ molecule in an optical cavity. (b) The Pauli-Fierz Hamiltonian that describes strong light-matter interactions. (c) The Pauli-Fierz Hamiltonian in the qubit basis along with the hybrid quantum-classical VQE algorithm for calculation of the ground state and the excited state energies employing the QED-UCC ansatz. (d) Simulated molecular properties such as ionization potentials, electron affinities, excitation energies, and absorption spectra in an optical cavity.}
  \label{fig:schematic_representation}
\end{figure*}

Motivated by these impressive performances of the UCCSD in VQE, and by the importance of the development of a reliable, robust, and scalable model in computational polaritonic chemistry, in this work we introduce the quantum electrodynamics unitary coupled cluster (QED-UCC) method as well as the quantum electrodynamics equation-of-motion (QED-EOM) method suitable for calculations on quantum computers. The herein developed methods are then used to study the strong light-matter interaction in an optical cavity in regimes where strong electronic correlations become significant. Schematic depiction of the workflow performed in this work is given in Fig. \ref{fig:schematic_representation}. The main purpose of this work is to provide the theoretical framework, the detailed working equations, and to discuss the performance of the aforementioned methods. To the best of our knowledge, this is the first UCC method in which both electrons and photons are treated quantum mechanically using the fully $\it ab$ $\it initio$ Hamiltonian. Moreover, this work lays the foundation for the development of other theoretical models in polaritonic chemistry suitable for quantum computers that are more robust in the regime of the strong electronic correlation~\cite{grimsley2019adaptive,lee2018generalized,ryabinkin2018qubit} than the UCC method. These developments can lead to further understanding of complex catalysts in an optical cavity. One such system is the iron molybdenum cofactor (FeMoco), the active site of the Nitrogenase enzyme that reduces molecular nitrogen at room temperature and standard pressure \cite{hoffman2014mechanism}. However this mechanism is not yet well understood due to the complex electronic structure of FeMoco. As discussed in Ref. \citenum{reiher2017elucidating}, quantum computer can be employed to better understand this reaction mechanism and an optical cavity may be used to provide an additional knob to suppress or enhance the catalytic activity of the aforementioned system or potentially providing more insight into the reaction mechanism.

\section{Theory}
The interaction of a molecule and the quantized electromagnetic field in an optical cavity with a single photon mode can be described by the Pauli-Fierz Hamiltonian~\cite{ruggenthaler2014quantum,tokatly2013time, ruggenthaler2018quantum} and formulated within the dipole approximation and in the length gauge reads as~\cite{deprince2021cavity,haugland2021intermolecular} 
\begin{equation}
\begin{aligned}
    \label{eqn:PF_Hamiltonian}
    \hat{H} &= h^p_q a_p^q + \frac{1}{2}g^{pq}_{rs} a_{pq}^{rs}+\omega_{\text{cav}}b^{\dagger}b\\&-\sqrt{\frac{\omega_\text{cav}}{2}}(\boldsymbol{\lambda} \cdot \boldsymbol{d})(b^{\dagger}+b)+\frac{1}{2}(\boldsymbol{\lambda} \cdot \boldsymbol{d})^2.
\end{aligned}
\end{equation}
While we consider for simplicity only the case of the single photon mode in this paper, the extension to multiple photon modes is straightforward~\cite{haugland2020coupled,white2020coupled}.
The first two terms constitute the electronic Hamiltonian (within the Born-Oppenheimer approximation), where $a_{p_1p_2...p_n}^{q_1q_2...q_n}=a_{q_1}^{\dagger}a_{q_2}^{\dagger}...a_{q_n}^{\dagger}a_{p_n}...a_{p_2}a_{p_1}$ are the second-quantized electronic excitation operators expressed as a string of fermionic creation ($a^{\dagger}$) and annihilation operators ($a$), $h^p_q=\langle q|\hat{h}^\text{e}|p\rangle$ is a matrix element of the core electronic Hamiltonian, and $g^{pq}_{rs}=\langle rs|pq\rangle$ is a two-electron repulsion tensor element in the physics notation. Here, $p,q,r,s,...$ indices denote general electronic spin orbitals, $i,j,k,l,...$ indices denote occupied electronic spin orbitals, and $a,b,c,d,...$ indices denote unoccupied electronic spin orbitals. The Einstein summation convention over repeated indices is invoked throughout this work. The third term denotes the oscillation of the cavity with frequency $\omega_\text{cav}$ that is expressed in terms of bosonic creation ($b^{\dagger}$) and annihilation ($b$) operators. The fourth term is the coupling between electronic and photonic degrees of freedom, where $\boldsymbol{\lambda}$ is the coupling strength vector, and $\boldsymbol{d}=\boldsymbol{d}_{\text{e}}+\boldsymbol{d}_{\text{nuc}}$ (electronic and nuclear, respectively) is the molecular dipole moment operator. Finally, the fifth term describes the dipole-self energy. This Hamiltonian can be conveniently transformed to the coherent state basis as discussed in Ref. \citenum{haugland2020coupled}, which we also adopt in this work.

In the QED-CC method~\cite{haugland2020coupled}, the wave function ansatz is given as
\begin{equation}
    \label{eqn:QED-CC}
    |\Psi_{\text{QED-CC}}\rangle=e^{\hat{T}}|0^{\text{e}}0^{\text{p}}\rangle,
\end{equation}
where $|0^{\text{e}}0^{\text{p}}\rangle=|0^{\text{e}}\rangle\otimes|0^{\text{p}}\rangle$ is the reference quantum electrodynamics Hartree-Fock (QED-HF) wave function expressed as a direct product between an electronic Slater determinant ($|0^{\text{e}}\rangle$) and a photon-number state ($|0^{\text{p}}\rangle$). Superscripts e and p denote electrons and photons, respectively. In this equation, the cluster operator
\begin{equation}
    \label{eqn:QED-T}
    \hat{T}=\hat{T}_1+\hat{T}_2+...=\sum_{\mu,n}t_{\mu,n}a^{\mu}(b^{\dagger})^n
\end{equation}
incorporates correlation effects between the quantum particles (i.e. electrons and photons) through single, double, and higher excitation ranks. Here, $a^\mu=a_\mu^{\dagger}=\{a_{i}^{a},a_{ij}^{ab},...\}$ is the electronic excitation operator, index $\mu$ is the electronic excitation manifold, and $n$ corresponds to the number of photons. The unknown $t_{\mu,n}$ parameters (amplitudes) are determined via the projective technique~\cite{bartlett2007coupled,shavitt2009many,haugland2020coupled} 
\begin{equation}
    \label{eqn:QED-T-equations}
    \langle0^{\text{e}}0^{\text{p}}|a_{\mu}(b)^ne^{-\hat{T}}\hat{H}e^{\hat{T}}|0^{\text{e}}0^{\text{p}}\rangle=\sigma_{\mu,n}.
\end{equation}
Additionally, the excitation energies are obtained from the QED-EOM-CC method~\cite{haugland2020coupled} by diagonalizing
\begin{equation}
    \label{eqn:QED-EOM-CCSD}
    J_{\mu m, \nu n} =\langle0^{\text{e}}0^{\text{p}}|a_{\mu}b^m[e^{-\hat{T}}\hat{H}e^{\hat{T}},a^{\nu}(b^{\dagger})^n]|0^{\text{e}}0^{\text{p}}\rangle.
\end{equation}

Truncation of the cluster operator to
\begin{equation}
    \label{eqn:QED-T-1}
    \hat{T}=t^{i,0}_a a_i^a+t^{0,1}b^\dagger+\frac{1}{4}t^{ij,0}_{ab} a_{ij}^{ab}+t^{i,1}_aa_i^ab^\dagger+\frac{1}{4}t^{ij,1}_{ab} a_{ij}^{ab}b^\dagger
\end{equation}
defines the QED-CCSD-1 method~\cite{haugland2020coupled} in which one photon interacts with up to two electrons. In addition to the cluster operator defined in Eq. \eqref{eqn:QED-T-1}, in this work we also consider 
\begin{equation}
\begin{aligned}
    \label{eqn:QED-T-2}
    \hat{T}&=t^{i,0}_a a_i^a+t^{0,1}b^\dagger+\frac{1}{4}t^{ij,0}_{ab} a_{ij}^{ab}+t^{i,1}_aa_i^ab^\dagger+\frac{1}{4}t^{ij,1}_{ab} a_{ij}^{ab}b^\dagger\\&+t^{0,2}b^\dagger b^\dagger+t^{i,2}_aa_i^ab^\dagger b^\dagger+\frac{1}{4}t^{ij,2}_{ab} a_{ij}^{ab}b^\dagger b^\dagger,
\end{aligned}
\end{equation}
where up to two photons interact with up to two electrons. We will refer to this method as QED-CCSD-2. Both of these methods have a computational cost that scales as $N^6$ on a classical computer, where $N$ is a measure of the system size. 

After setting up the QED-CC method, we now define the unitary coupled cluster method for light-matter systems, the QED-UCC method. In the QED-UCC method, the wave function ansatz is given by
\begin{equation}
    \label{eqn:QED-UCC}
    |\Psi_{\text{QED-UCC}}\rangle=e^{\hat{T}-\hat{T}^\dagger}|0^{\text{e}}0^{\text{p}}\rangle.
\end{equation}
In this method, the unknown amplitudes $t_{\mu,n}$ are determined by variational optimization of the following energy functional 
\begin{equation}
    \label{eqn:QED-UCC-Funtional}
     E_{\text{QED-UCC}}=\underset{t}{\text{min}} \langle\Psi_{\text{QED-UCC}}|\hat{H}|\Psi_{\text{QED-UCC}}\rangle
\end{equation}
that provides the QED-UCC energy using the Hamiltonian of Eq. \eqref{eqn:PF_Hamiltonian}. In here, we consider the two different methods, namely QED-UCCSD-1 and QED-UCCSD-2, with the cluster operator $\hat{T}$ defined in Eqs. \eqref{eqn:QED-T-1} and \eqref{eqn:QED-T-2}. These two methods, even when $\hat{T}$ is truncated, have exponential cost on a classical computer due to nontruncation of the Baker-Campbell-Hausdorff expansion~\cite{taube2006new}. In addition to these QED-UCC methods, we also investigate the generalized unitary coupled cluster method within the QED framework where the electronic excitation operators in Eqs. \eqref{eqn:QED-T-1} and \eqref{eqn:QED-T-2} are replaced with $\{a_{p}^{q},a_{pq}^{rs}\}$~\cite{nooijen2000can,lee2018generalized}. These generalized QED-UCC methods are denoted as QED-GUCCSD-1 and QED-GUCCSD-2. All of these methods are both variational and unitary. The later property is particularly important because quantum computers implement unitary operations. However, a practical implementation of the UCC methods on current NISQ devices requires that the ansatz given with Eq. \eqref{eqn:QED-UCC} is approximated by the first-order Trotter-Suzuki expansion as
\begin{equation}
    \label{eqn:QED-Trotter}
    e^{\sum_{\mu,n}t_{\mu,n}\big[a^{\mu}(b^{\dagger})^n-a_{\mu}b^n\big]}\approx\prod_{\mu,n} e^{t_{\mu,n}\big[a^{\mu}(b^{\dagger})^n-a_{\mu}b^n\big]},
\end{equation}
which is one cause for errors~\cite{romero2018strategies,grimsley2019trotterized,evangelista2019exact,izmaylov2020order,childs2021theory} in the calculation. Analogously to the electronic case, the solution of the truncated and Trotterized QED-UCC method in conjunction with the VQE algorithm can be achieved at the polynomial cost on a quantum computer~\cite{peruzzo2014variational}.

In addition to the ground state energies, the VQE algorithm has been extended for the calculation of excitation energies on quantum computers~\cite{mcclean2017hybrid,lee2018generalized}. One such approach is the EOM method~\cite{rowe1968equations,ollitrault2020quantum} where the excitation energy of the $k$th excited state is calculated from
\begin{equation}
    \label{eqn:QED-EOM-Excitation}
    \Delta E_k=\frac{\langle\Psi_{\text{QED-UCC}}|[O_k,\hat{H},O_k^{\dagger}]|\Psi_{\text{QED-UCC}}\rangle}{\langle\Psi_{\text{QED-UCC}}|[O_k,O_k^{\dagger}]|\Psi_{\text{QED-UCC}}\rangle},
\end{equation}
where $O_k^{\dagger}=\sum_{\mu,n}\big[X_{\mu,n}(k)a^{\mu}(b^{\dagger})^n-Y^{\mu,n}(k)a_{\mu}b^n\big]$ is the excitation operator and commutators are $[O_k,O_k^{\dagger}]=O_kO_k^{\dagger}-O_k^{\dagger}O_k$ and $[O_k,\hat{H},O_k^{\dagger}]=([[O_k,\hat{H}],O_k^{\dagger}]+[O_k,[\hat{H},O_k^{\dagger}]])/2$. Diagonalization of the generalized eigenvalue equation
\begin{equation}
    \label{eqn:QED-EOM-Equation}
    \begin{pmatrix}
        \mathbf{A} & \mathbf{B}\\
        \mathbf{B}^* & \mathbf{A}^*
    \end{pmatrix}
    \begin{pmatrix}
        \mathbf{X}(k)\\
        \mathbf{Y}(k)
    \end{pmatrix}
    =\Delta E_k
    \begin{pmatrix}
        \mathbf{C} & \mathbf{D}\\
        -\mathbf{D}^* & -\mathbf{C}^*
    \end{pmatrix}
    \begin{pmatrix}
        \mathbf{X}(k)\\
        \mathbf{Y}(k)
    \end{pmatrix},
\end{equation}
where matrix elements are defined as
\begin{equation}
    \label{eqn:QED-EOM-Mat-Elem}
        \begin{aligned}
            A_{\mu m,\nu n}&=\langle\Psi_{\text{QED-UCC}}|[a_{\mu}b^m,\hat{H},a^{\nu}(b^{\dagger})^n]|\Psi_{\text{QED-UCC}}\rangle, \newline\\
            A_{\mu m,\nu n}&=-\langle\Psi_{\text{QED-UCC}}|[a_{\mu}b^m,\hat{H},a_{\nu}b^n]|\Psi_{\text{QED-UCC}}\rangle, \newline\\
            A_{\mu m,\nu n}&=\langle\Psi_{\text{QED-UCC}}|[a_{\mu}b^m,a^{\nu}(b^{\dagger})^n]|\Psi_{\text{QED-UCC}}\rangle, \newline\\
            A_{\mu m,\nu n}&=-\langle\Psi_{\text{QED-UCC}}|[a_{\mu}b^m,a_{\nu}b^n]|\Psi_{\text{QED-UCC}}\rangle, 
        \end{aligned}
\end{equation}
provides the excitation energy as well as the unknown excitation amplitudes $X$ and $Y$.

The quantum algorithms in this work follow the usual steps of the VQE algorithm along with additional modifications due to the quantum treatment of the bosonic particles (photons). We provide the procedure in the following: In the first step, the second-quantized Hamiltonian defined in Eq. \eqref{eqn:PF_Hamiltonian} is constructed using the spin-orbitals obtained from the QED-HF calculation. The second step requires the mapping of the spin-orbitals to the qubit basis. Because electrons and photons are governed by different spin statistics, different mappings need to be utilized for different types of particles. Fermionic creation and annihilation operators obey the usual anticommutation relations $(\{a_p,a_q\}=\{a_p^{\dagger},a_q^{\dagger}\}=0$, and $\{a_p^{\dagger},a_q\}=\delta_{pq})$, bosonic creation and annihilation operators obey the usual commutation relations $([b_p,b_q]=[b_p^{\dagger},b_q^{\dagger}]=0$, and $[b_p^{\dagger},b_q]=\delta_{pq})$, and the operators between particles of different types commute $([a_p,b_q]=[a_p^{\dagger},b_q^{\dagger}]=[a_p^{\dagger},b_q]=[a_p,b_q^{\dagger}]=0)$.
For the case of the electronic operators, we map the $p$th electronic spin-orbital to the $p$th qubit via the Jordan-Wigner transform~\cite{jordan1993collected} where $n$ spin-orbitals are mapped to $n$ qubits. In this work, we are using OpenFermion's~\cite{mcclean2020openfermion} implementation of the Jordan-Wigner transform. In this transformation, the fermionic creation operator that creates the $p$th electronic spin-orbital is expressed in the basis of the Pauli matrices ($\sigma_x,\sigma_y,\sigma_z,I$) as $a^{\dagger}_p=1/2(\sigma_x^p-i\sigma_y^p)\otimes_{q<p}\sigma_z^q$. In contrast, for the case of bosons (photons), we use the direct boson mapping~\cite{somma2003quantum,veis2016quantum}, where the $p$th photonic spin-orbital with $n$ photons is mapped to $n+1$ qubits. In this transformation, the bosonic creation operator is given by
\begin{equation}
    \label{eqn:Boson-mapping}
        \begin{aligned}
            b^{\dagger}_p=\sum_{j=0}^{n-1}(j+1)\sigma_{+}^{j,p}\sigma_{-}^{j+1,p},
        \end{aligned}
\end{equation}
where $\sigma_{\pm}=1/2(\sigma_x\pm i\sigma_y)$, and the number operator is given by
\begin{equation}
    \label{eqn:Number-operator-mapping}
        \begin{aligned}
            b^{\dagger}_p b_p=\sum_{j=0}^{n}j \frac{\sigma_z^{j,p}+I}{2}.
        \end{aligned}
\end{equation}
 
Using these two steps, we convert the Hamiltonian of Eq.~\eqref{eqn:PF_Hamiltonian} to the qubit basis $\hat{H}=\sum_i c_i(\mathbf{R})\hat{P}_i$, where $c_i(\mathbf{R})$ are the Hamiltonian parameters obtained in the first step that depend on the molecular geometry, and $\hat{P}_i$ are products of the Pauli matrices. Additionally, the cluster operators in Eqs. \eqref{eqn:QED-T-1}, \eqref{eqn:QED-T-2}, and \eqref{eqn:QED-EOM-Mat-Elem} are converted to the qubit basis using the same transformation scheme described in the second step.
After these two steps, we can minimize the energy functional. Here, the energy functional in Eq. \eqref{eqn:QED-UCC-Funtional} is variationally minimized with the SciPy's implementation of the Broyden-Fletcher-Goldfarb-Shanno algorithm~\cite{2020SciPy}. This step provides the optimal amplitudes $t$ in addition to the ground state QED-UCC energy. In the last step, we use these optimal amplitudes for calculation of the excitation energy by diagonalizing Eq. \eqref{eqn:QED-EOM-Equation}. 

All discussed methods were implemented into an in-house developmental version of the Psi4NumPy quantum chemistry software~\cite{smith2018psi4numpy}, which will be made publicly available in near future. The QED-CCSD-$n$, where $n=1,2$ indicate number of photons, and the QED-EOM-CCSD-1 methods are implemented in the traditional way suitable for classical computers~\cite{bartlett2007coupled,haugland2020coupled}, whereas the QED-FCI-$n$, QED-UCCSD-$n$, QED-GUCCSD-$n$, and EOM-QED-UCCSD-$1$ methods are implemented in the qubit basis to be suitable for implementations on quantum computers. The implementation was validated using the H$_2$ molecule. Because this system contains only two electrons, all the implemented methods are exact for a given basis set and will give the same result for the same number of photons. Details of these benchmark calculations are provided in the Section 1 of the Supplementary Information.

\section{Results}
We now focus our attention to a system in which the strong electronic correlation is significant and for which the truncated methods are not exact. Note that the strong electronic correlation is important in regimes where the wave function is not described qualitatively well by a single Slater determinant~\cite{lyakh2012multireference}. For that purpose, we compute the potential energy surface of the H$_4$ model with geometry 
\begin{equation}
    \label{eqn:H4-geometry}
        \begin{aligned}
            &\text{H1}: [0, 0, 0] \text{\AA} \\\newline\nonumber
            &\text{H2}: [0, 0, 1.23] \text{\AA} \\\newline\nonumber
            &\text{H3}: [R, 0, 0] \text{\AA} \\\newline\nonumber
            &\text{H4}: [R, 0, 1.23] \text{\AA} \nonumber
        \end{aligned}
\end{equation}
as defined in Ref. \citenum{lee2018generalized}. This model is an excellent system for testing the validity of various methods when the strong electronic correlation is significant~\cite{evangelista2006high,small2012fusion,lee2018generalized}. For $R=1.23 \text{\AA}$ (geometry with the $D_{4h}$ symmetry), two reference Slater determinants become quasidegenerate in energy, which is challenging for the CCSD method as shown previously~\cite{small2012fusion,lee2018generalized}. 

Here we test the performance of the proposed QED methods with one and two photons by calculating ionization potentials (IP) and electron affinities (EA) in an optical cavity when $R$ is varied. Accurate predictions of IP and EA is essential for calculation of redox potentials, band gaps, and for a better understanding and design of photovoltaic materials~\cite{gratzel2003solar}. Many of these materials can exhibit strong electronic correlation character that is posing a challenge for the single reference methods. Therefore, development of methods for reliable predictions in this regime is an active field of research~\cite{chatterjee2019second}. A recent QED-CCSD-1 study indicated that the IP and EA of sodium halide compounds can be significantly modified in the presence of an optical cavity~\cite{deprince2021cavity}. IP and EA are defined as 
\begin{equation}
    \label{eqn:IP_EA}
        \begin{aligned}
            E_{\text{IP}} &= E_+ - E_0,\\\newline
            E_{\text{EA}} &= E_0 - E_-,
        \end{aligned}
\end{equation}
where $E_0$, $E_+$, and $E_-$ are energies of neutral, cationic, and anionic states, respectively. 

\begin{figure*}[ht!]
  \centering
  \includegraphics[width=6.5in]{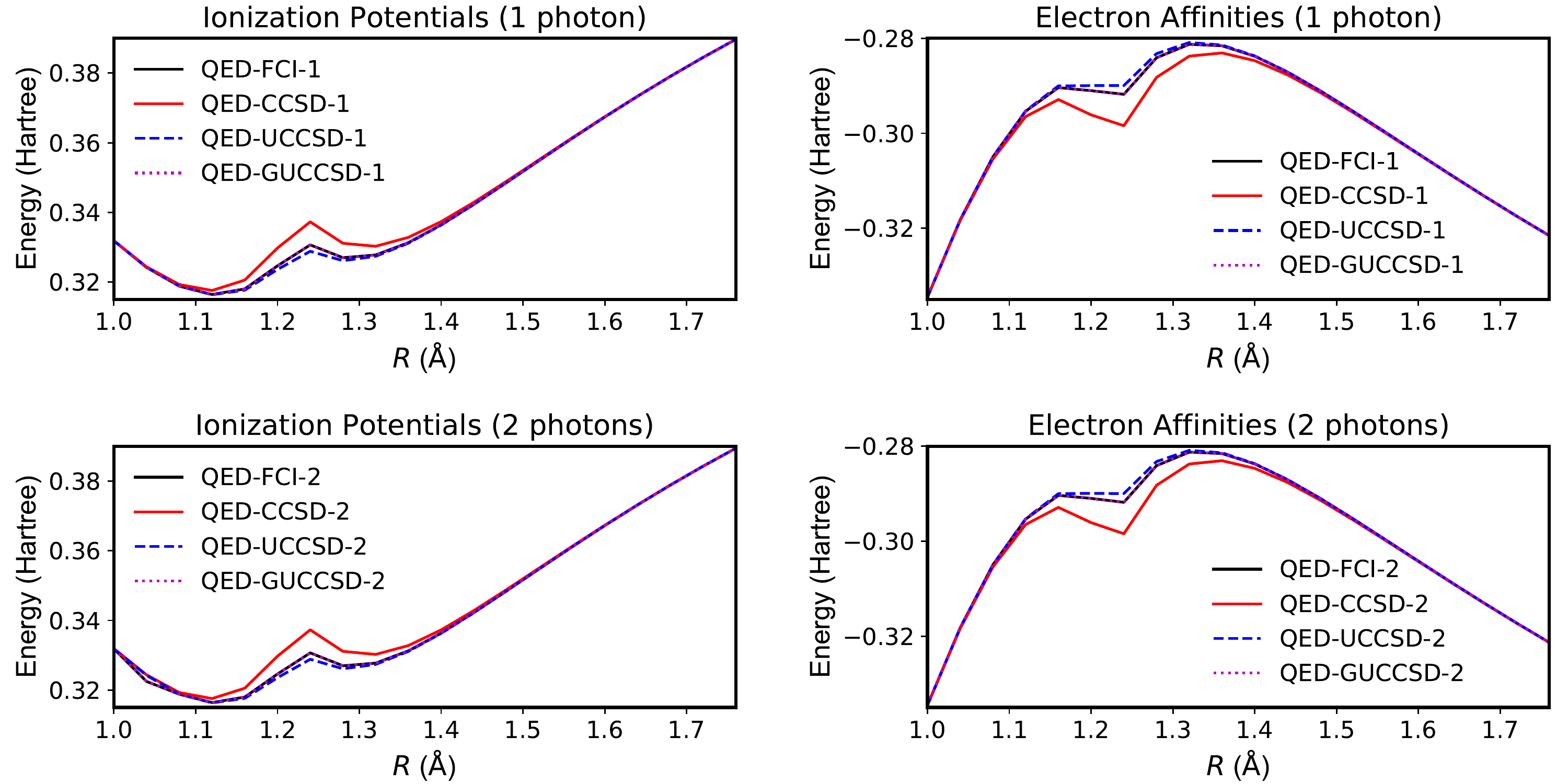}
  \caption{Calculated vertical IP (left panels) and EA (right panels) values with the various QED methods for one photon (top panels) and two photons (bottom panels). The calculations employ the STO-3G basis set with the cavity parameters $\omega_{\text{cav}}=10$ eV and $\boldsymbol{\lambda}=[0.1, 0, 0]$ a.u.}
  \label{fig:ip_ea}
\end{figure*}

Figure \ref{fig:ip_ea} shows vertical IP (left panels) and EA (right panels) calculated with the different QED methods including the one photon state (top panels) and two photon states (bottom panels) for the H$_4$ molecule inside an optical cavity as the distance $R$ is increased. This calculation is performed in an optical cavity of a cavity mode with the frequency $\omega_\text{cav}=10$ eV, the coupling strength vector $\boldsymbol{\lambda}=[0.1, 0, 0]$ in atomic units, and by employing the STO-3G basis set. The quantum algorithms for this system and the STO-3G basis set in the case of one or two photons require 10 qubits or 11 qubits in total, respectively, out of which 8 qubits are required for the 8 electronic spin-orbitals. The reference values are calculated with the QED-FCI-1 (top panels black solid) and QED-FCI-2 methods (bottom panels black solid). Both IP and EA calculated with the QED-CCSD-1 (top panels red solid) and QED-CCSD-2 (bottom panels red solid) methods show the largest deviations relative to the reference values around the $D_{4h}$ geometry ($1.1\text{\AA}<R<1.4\text{\AA}$) where the strong electronic correlation is pronounced. Outside this region, both methods are in an excellent agreement with respect to the reference values. Note that the QED-CCSD-1 and QED-CCSD-2 methods are nonvariational and this behaviour is shown in Supplementary Fig. 1. The QED-UCCSD-1 (top panels blue dashed) and QED-UCCSD-2 (bottom panels blue dashed) methods display much lower discrepancy and are in a good agreement with the reference values even for the regions where the strong electronic correlation effects are pronounced. The QED-UCCSD-1 and QED-UCCSD-2 methods are variational and remain upper bound with respect to the reference values for all $R$ as shown in Supplementary Fig. 1. Lastly, the QED-GUCCSD-1 (top panels mangenta dotted) and QED-GUCCSD-2 (bottom panels mangenta dotted) methods are numerically exact and overlaps with the reference curves. A similar behaviour has been observed in the equivalent conventional electronic structure methods for the total energies~\cite{lee2018generalized}. As observed previously~\cite{deprince2021cavity}, EA are more sensitive than the IP in the presence of an optical cavity, and in this work we observe the same behaviour. The change in EA of the H$_4$ molecule inside the cavity with respect to outside the cavity is $\sim$0.15 eV on average, whereas IP is $\sim$0.03 eV on average, although a meaningful conclusion would require much larger basis sets which is beyond the scope of this work.

As visible from Fig. \ref{fig:ip_ea}, addition of the second photon states has a negligible effect on the calculated IP and EA values for this calculation setup. The effect of the second photon becomes evident for larger values of the coupling strength as shown in Supplementary Fig. 2, where the same quantities are calculated with the $\boldsymbol{\lambda}=[0.3, 0, 0]$ a.u. Addition of the second photon changes the resulting IP and EA on average by $\sim$0.15 eV. This result shows that addition of the second photon further enhances the effect of an optical cavity resulting in an additional changes of molecular properties. 

\begin{figure}[ht]
  \centering
  \includegraphics[width=3.25in]{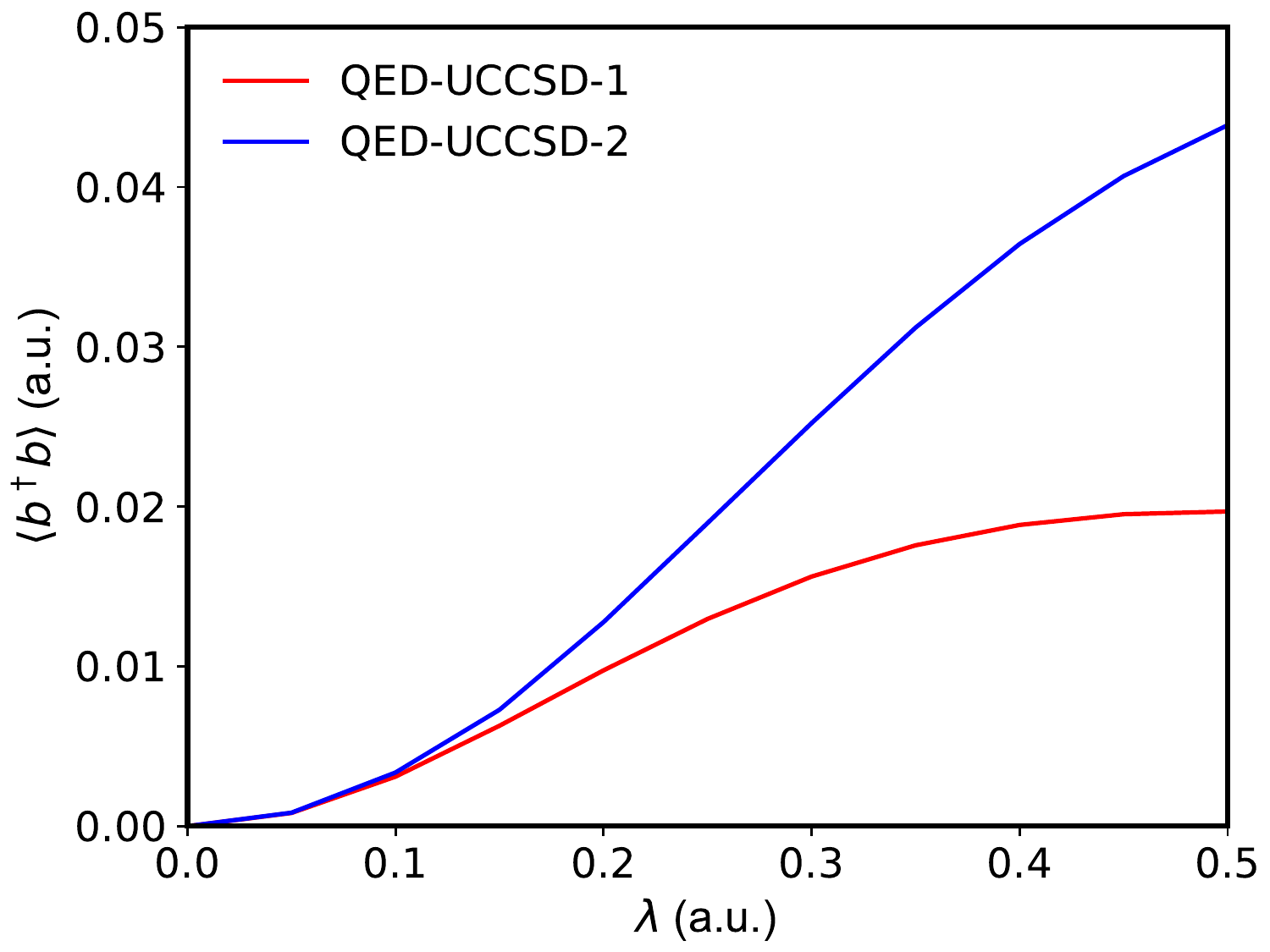}
  \caption{Expectation value of the photon occupation number operator $\langle b^\dagger b \rangle$ as a function of the coupling strength $\boldsymbol{\lambda}=[\lambda, 0, 0]$ a.u. for the QED-UCCSD-1 (red solid) and QED-UCCSD-2 (blue solid) methods. The calculations were performed with the fixed geometry parameter $R=1 \text{\AA}$ and  $\omega_{\text{cav}}=10$ eV.}
  \label{fig:number_operator}
\end{figure}

To gain a deeper insight between the methods that can treat up to one and two photons, we have calculated the expectation value of the photon occupation number operator ($\langle b^\dagger b \rangle$) with the QED-UCCSD-1 (red) and QED-UCCSD-2 (blue) methods, and its dependence on the coupling strength $\lambda$ is shown in Fig. \ref{fig:number_operator}. For the small values of coupling strengths ($\lambda < 0.15$ a.u.), the two methods predicts a similar values for this quantity. For larger values of coupling strength ($\lambda > 0.15$  a.u.), the computed quantity differs significantly between these two methods, indicating that the two photon processes are more pronounced at larger coupling strengths consistent with findings in previous work~\cite{flick2018abinito}. This is in agreement with our statement and observation from the previous paragraph. Lastly, the same quantity is computed with the QED-GUCCSD-1 and QED-GUCCSD-2 methods and the results are nearly identical to those obtained with the QED-UCCSD-1 and QED-UCCSD-2 methods, respectively.

The energy errors of a given method relative to the reference values increase with an increase of the coupling strength $\lambda$ as shown in Supplementary Fig. 3. Additionally, Supplementary Fig. 3 shows that this error is more pronounced for geometries where the strong electronic correlation is significant. We can understand this property with the mixing in of bare excited states into the correlated ground state for higher coupling strength~\cite{flick2018abinito,PhysRevA.98.043801}. Errors due to Trotterization for discussed methods are shown in Supplementary Fig. 4. 

Next, we investigate the performance of the developed methods for calculation of the excitation energies.
\begin{figure}[ht]
  \centering
  \includegraphics[width=3.25in]{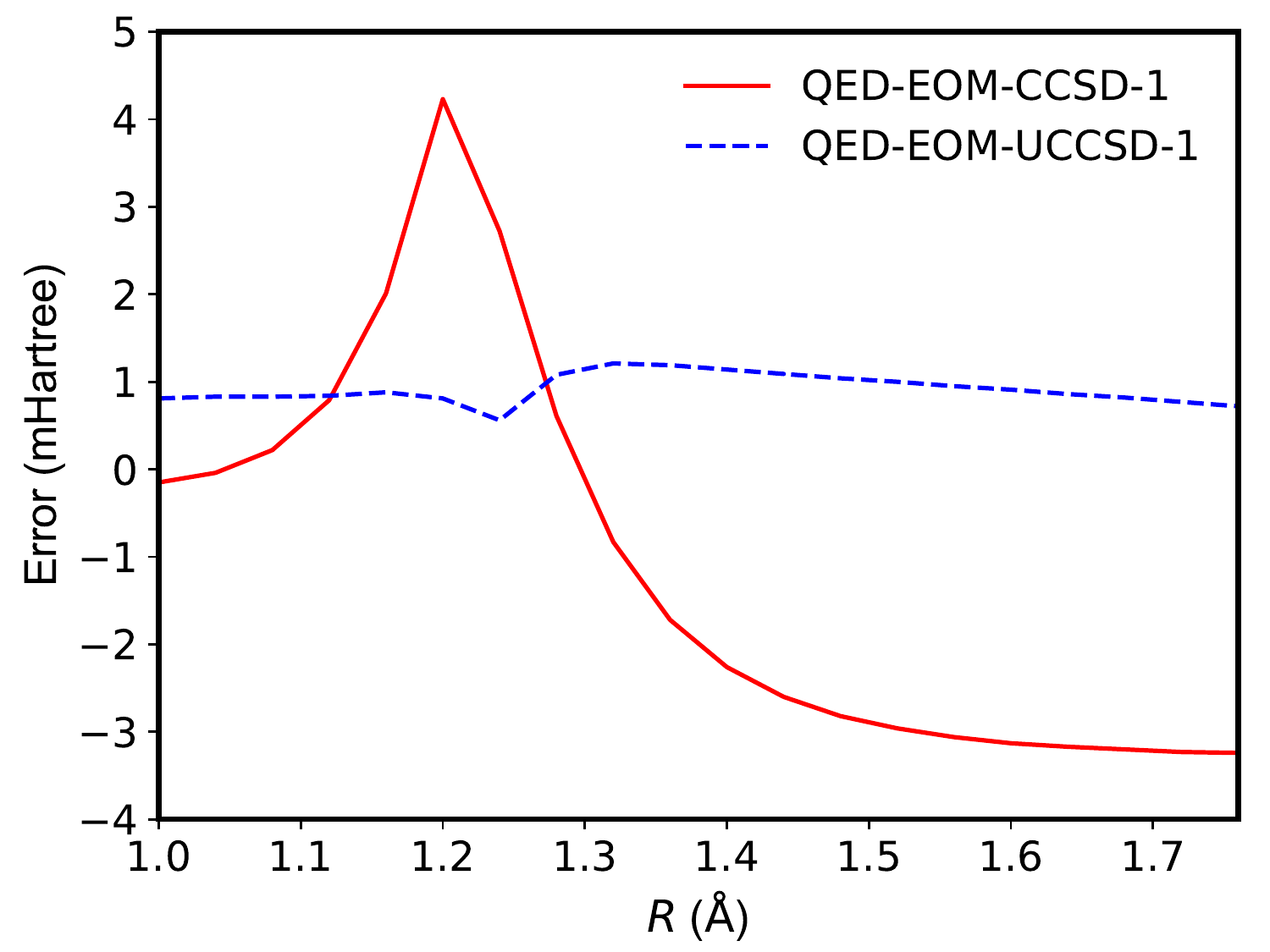}
  \caption{Errors in energies for the first excited state calculated with the QED-EOM-CCSD-1 (red solid) and the QED-EOM-UCCSD-1 (blue dashed) methods relative to the QED-FCI-1 method. The calculations were performed by employing the STO-3G basis set with the cavity parameters $\omega_{\text{cav}}=10$ eV and $\boldsymbol{\lambda}=[0.1, 0, 0]$ a.u.}
  \label{fig:error_excited}
\end{figure}
Figure \ref{fig:error_excited} shows the errors (in mHartree) in excitation energies for the first excited state calculated with the QED-EOM-CCSD-1 and the QED-EOM-UCCSD-1 methods. All errors are relative to the QED-FCI-1 method. This excited state corresponds to electronic spin-singlet excited state with the predominant single electronic excitation. In the case of the QED-EOM-UCCSD-1, the character of the excitation is determined from the leading $X$ amplitudes obtained by diagonalizing Eq. \eqref{eqn:QED-EOM-Equation}, whereas in the case of QED-EOM-CCSD-1, the character is determined by analysing the equivalent quantity after diagonalizing Eq. \eqref{eqn:QED-EOM-CCSD}. It is evident from this Fig. that the QED-EOM-UCCSD-1 method exhibits much smaller errors and deviations than the QED-EOM-CCSD-1 method. A similar behaviour is observed for the system outside an optical cavity. 

\begin{figure}[ht]
  \centering
  \includegraphics[width=3.25in]{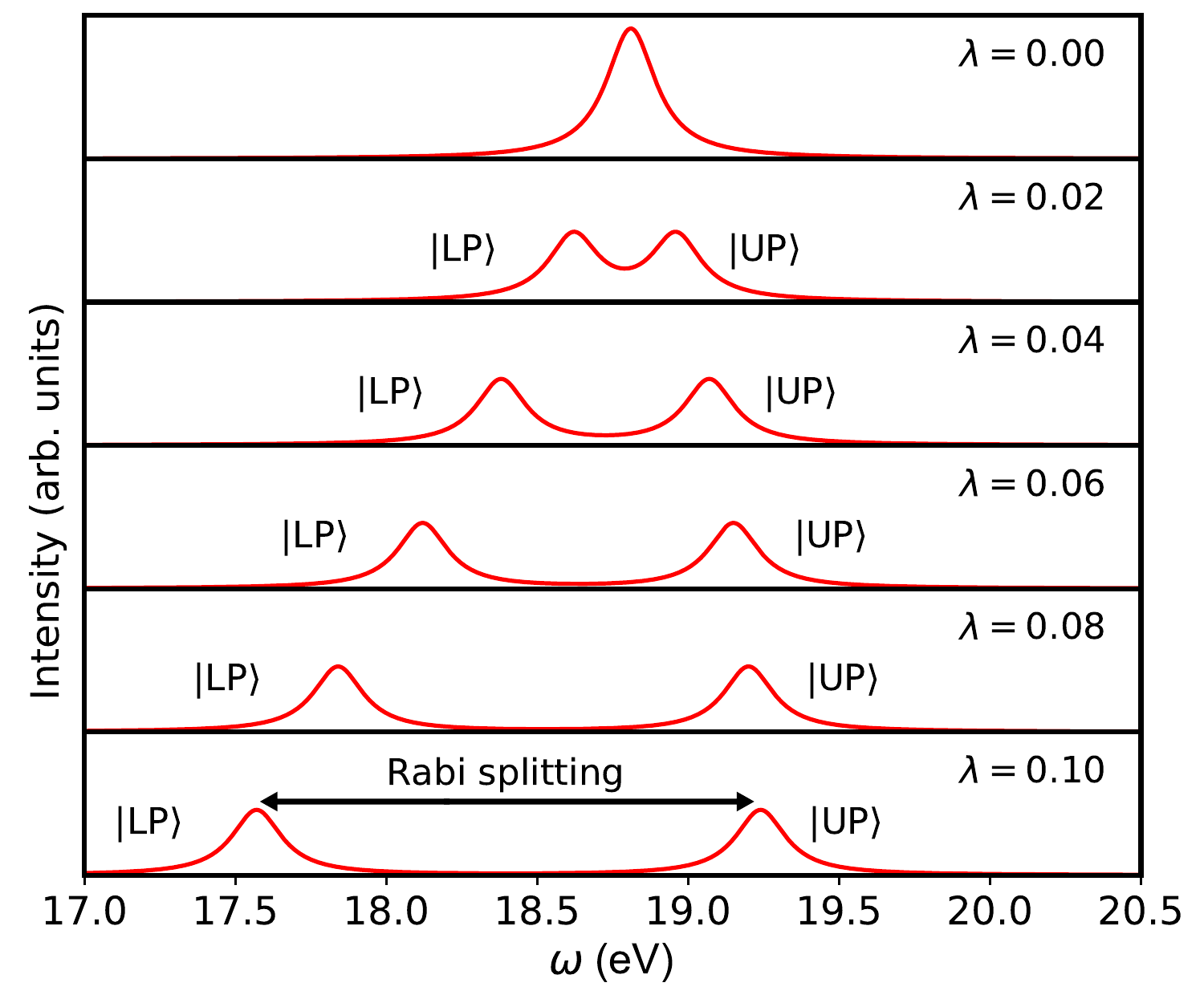}
  \caption{Absorption spectra of the H$_4$ molecule calculated with the QED-EOM-UCCSD-1 method for different coupling strengths $\lambda$. The calculations were performed with the fixed geometry parameter $R=1 \text{\AA}$ and $\omega_{\text{cav}}=18.81$ eV. The spectra is simulated by employing the Lorentzian broadening ($\Delta=0.01$ eV) and by assuming that both peaks have the same values of the oscillator strength.}
  \label{fig:rabi_splitting}
\end{figure}

Finally, we compute the Rabi splitting with the QED-EOM-UCCSD-1 method between two polariton states, $|\text{LP}\rangle$ and $|\text{UP}\rangle$, for the radiation field that is in resonance with the second excited state. The excitation energy corresponding to this state calculated with the EOM-UCCSD method is 18.81 eV with transition dipole moment along the x-axis. Figure \ref{fig:rabi_splitting} shows the non-symmetric Rabi splitting as the coupling strength vector $\boldsymbol{\lambda}=[\lambda, 0, 0]$ increases. For the coupling strength parameter $\lambda=0.1$, the upper polariton increases by 0.43 eV, where as lower polariton decreases by 1.24 eV. This non-symmetric Rabi splitting is caused by the effective detuning due to the $(\boldsymbol\lambda\cdot \boldsymbol d)^2$ term in Eq.~\eqref{eqn:PF_Hamiltonian}~\cite{doi:10.1021/acsphotonics.9b00768}.  The exact QED-FCI-1 method predicts the same values of the Rabi splittings. 

\section{Summary and Conclusion}
In this work, we introduce different QED-UCC methods and the QED-EOM-UCC method for the ground- and excited-state energy calculations suitable for calculations on quantum devices. Although, the discussion in this work focuses on the regime of strong light-matter interactions, the methodology presented here is identical for the situation where electrons and harmonic phonons are treated quantum mechanically~\cite{white2020coupled}. The performance of the developed methods is tested on the H$_4$ system that exhibits strong electronic correlation character. The results are then compared with the exact values as well as to previously developed methods, such as the QED-CCSD-1 method~\cite{haugland2020coupled}, that are designed for calculations on a classical computer. To the best of our knowledge these calculation include for the first time explicit effects of two photon correlations within the first principle methods for polaritonic chemistry. Our results obtained for both the ground- and excited-state properties indicate that the developed methods are in an excellent agreement with the exact reference values and that they outperform their traditional counterparts in the regions where the strong electronic correlation is significant. Future work will include the application of this framework on existing NISQ devices, such as the IBM Q framework~\cite{cross2017open}. The formalism presented in this work opens up many additional research directions for developments in both the computational polaritonic quantum chemistry as well as in utilizing quantum devices most efficiently for strongly coupled electron-boson systems.


\section*{Acknowledgments}
The Flatiron Institute is a division of the Simons Foundation.

\section*{Author contributions}
F.P. conceived the project, wrote the code, obtained and evaluated the data, prepared the figures and the first draft. Both authors were involved in analysing and discussing the results, and editing the manuscript.\newline\\





\linespread{1}\selectfont
\bibliography{Journal_Short_Name,references}

\end{document}